\documentclass{emulateapj}
\usepackage[usenames,dvipsnames]{xcolor}
\definecolor{blue}{rgb}{0,0,1}
\usepackage{amsmath,amssymb,amsthm}
\usepackage{natbib}
\usepackage{graphicx}
\usepackage{float}
\usepackage[caption=false]{subfig}

\usepackage{amsmath}
\usepackage{amssymb}

\usepackage[hyperfootnotes=false]{hyperref}

\hypersetup{
  colorlinks,
  citecolor=blue,
 linkcolor=blue,
 urlcolor=Green}

\slugcomment{}

\shortauthors{Zhang et al. 2014}

\begin{document}

\title{Migration and Growth of Protoplanetary Embryos I: 
Convergence of Embryos in Protoplanetary Disks}

\author{Xiaojia Zhang\altaffilmark{1}$^*$, Beibei Liu\altaffilmark{2}, 
Douglas N. C. Lin\altaffilmark{1,3},
Hui Li,\altaffilmark{4}}
\altaffiltext{*}{E-mail: \href{mailto:xiaojia.f.zhang@gmail.com}
{xzhang47@ucsc.edu}}
\altaffiltext{1}{Department of Astronomy and Astrophysics, University 
of California, Santa Cruz, CA 95064, USA}
\altaffiltext{2}{Kavli Institute for Astronomy \& Astrophysics and 
Department of Astronomy, School of Physics, Peking University, Beijing 
100871, China}
\altaffiltext{3}{Institute for Advanced Studies, Tsinghua University, Beijing
100084, China}
\altaffiltext{4}{Los Alamos National Laboratory, Los Alamos, NM 87545, USA}

\begin{abstract}


According to the core-accretion scenario, planets form in protostellar disks 
through the condensation of dust, coagulation of planetesimals, and emergence 
of protoplanetary embryos.  At a few AU in a minimum mass nebula, embryos' 
growth is quenched by dynamical isolation due to the depletion of 
planetesimals in their feeding zone. However, embryos with masses ($M_p$) in 
the range of a few Earth masses ($M_\oplus$) migrate toward a transition 
radius between the inner viscously heated and outer irradiated regions of 
their natal disk. Their limiting isolation mass increases with the 
planetesimals surface density. When $M_p > 10 M_\oplus$, embryos 
efficiently accrete gas and evolve into cores of gas giants.  
We use numerical simulation to show that, despite 
streamline interference, convergent embryos essentially retain the
strength of non-interacting embryos' Lindblad and corotation torque 
by their natal disks. In disks with modest surface density (or 
equivalently accretion rates), embryos capture each other in their 
mutual mean motion resonances and form a convoy of super Earths.  
In more massive disks, they could overcome these resonant barriers to undergo 
repeated close encounters including cohesive collisions which enable 
the formation of massive cores.  
\end{abstract}


\section{Introduction}


Since the discovery of 51~Pegasi~b \citep{1995Natur.378..355M}, 
more than $10^3$ planets orbiting stars other than the Sun have been 
detected and confirmed. Exoplanet observations implicate that 
nearly $15-20\%$ of solar type stars harbor at least one gas 
giant planet with mass ($M_p$) comparable or larger than that 
of Saturn \citep{2008PhST..130a4001M,2008PASP..120..531C}. Despite the
observational bias against the detection of additional long-period 
and low-mass companions, a large fraction of the known gas giants 
reside in multiple-planet systems (http://exoplanet.eu/catalog-all.php). 
Both radial velocity and transit surveys lead to the discovery of
a much richer population of super Earths
 \citep{2011ESS.....2.0102M, 2012ApJS..201...15H, 2013ApJS..204...24B, 2013ApJ...766...81F, 2013ApJ...770...69P}. 
These findings pose a constraint on the theory of planet formation.

The widely adopted core accretion scenario is based on the assumption that 
planet formation proceeded through dust condensation and aggregation, 
planetesimal coagulation, and embryo mergers within their gaseous natal 
disks \citep{1993ARA&A..31..129L, 2004ARA&A..42..549G, 2005A&A...434..971D}. 
This growth is quenched when the embryos acquire an isolation mass 
($M_{\rm iso}$) through the consumption of most planetesimals within 
their feeding zones \citep{1987Icar...69..249L,1998Icar..131..171K}. 
At a few AU in a minimum mass nebula (MMN), $M_{\rm iso} \sim$ is a few Earth
mass ($M_\oplus$) but below the critical mass $M_{\rm crit} (\sim 10
M_\oplus)$ \citep{1996Icar..124...62P}. Although $M_{\rm iso} >
M_{\rm crit}$ well beyond the snow line, embryos growth time scale
in the outer disk regions is likely to exceed the observed disk depletion
time scale ($\tau_{\rm dep} \sim 3-5$ Myr)  \citep{2004ApJ...604..388I}.

The magnitude of $M_{\rm iso} \propto \Sigma_p ^{3/2}$ where $\Sigma_p$
is the surface density of the building block planetesimals.  The growth 
barrier can be bypassed with either an initial $\Sigma_p$ a few times
larger than that of the MMN model or new supplies of grains, planetesimals, 
or embryos to replenish the feeding zones. Potential transport mechanisms
include hydrodynamic drag of grains \citep{2003Icar..165..438W} and vortice-trapping 
of pebbles \citep{2010MNRAS.404..475J}. In this paper, we focus our discussion on the 
type I migration of embryos as the dominant mechanism to redistribute embryos' 
building block materia l\citep{2005A&A...434..343A}.

Embryos excite density waves in the disk which carries flux of angular 
momentum \citep{1982ARA&A..20..249G}. When these waves are dissipated,
they induce both differential Lindblad and corotation torque. Linear 
calculations indicate that although Earth-mass embryos do not 
significantly modify the disk structure, their angular momentum exchange
with the disk can lead to rapid type I migration \citep{1997Icar..126..261W, 2002ApJ...565.1257T, 
2012ARA&A..50..211K, 2013arXiv1312.4293B}.
At 5 AU in a MMN (where Jupiter resides today), differential Lindblad torque
alone leads to the inward migration of critical mass cores over a time scale 
$\tau_{\rm mig} < \tau_{\rm dep}$.  Such a rapid migration rate would not
only reduce the retention efficiency of cores but also suppress the formation
of gas giants at a few AU \citep{2005A&A...433..247P}.

Several mechanisms to reduce the rate of type I migration have been suggested.
They include stochastic migration in turbulent disks \citep{2005fdda.conf...29P}, 
"planet-trapping" by disk region with positive surface density 
gradient \citep{2006ApJ...642..478M, 2008A&A...478..929M} or by 
vortices generated
from Rossby wave instabilities \citep{2003ApJ...596L..91K, 2009ApJ...690L..52L}, and stalling migration 
at the inner disk edge \citep{2007ApJ...654.1110T, 2008A&A...478..939P}.

The most promising mechanism to quench type I migration is contribution 
by a corotation torque due to an embryo's interaction with nearby gas which 
follows a horseshoe pattern in its corotating frame. The magnitude
and sign of the corotation torque are determined by either the gas surface 
density ($\Sigma_g$) or entropy ($S_g$) gradient in the horseshoe region
 \citep{2006A&A...459L..17P, 2008A&A...487L...9K}. In the viscously 
heated inner and irradiated outer unperturbed regions of the disk, 
embryos' corotation 
torque is more intense than their differential Lindblad torque and has 
respectively positive and negative signs with the potential to induce 
their outward/inward migration \citep{2012ApJ...755...74K}.  
 
However, relatively massive embryos significantly perturb the disk stream 
lines, reduce their advective transport of angular momentum, and trap them 
within the horseshoe region (with a width $\Delta a_{\rm hr}$).  On a 
liberation time scale ($\tau_{\rm lib}$), mixing of the gas reduces the 
$\Sigma_g$ or $S_g$ gradients and weakens (saturates) the corotation torque
 \citep{2001MNRAS.326..833B}.  Nevertheless, turbulence in the disk also
induces an intrinsic outward transport of angular momentum which 
enables gas 1) to flow through the horseshoe region of embryos with modest 
masses, 2) to preserve the $\Sigma_g$ and $S_g$ distribution, and 3) to 
retain the corotation torque \citep{2010MNRAS.401.1950P, 2010ApJ...709..759B, 
2011MNRAS.410..293P}. Around low-mass embryos, the time scale for the 
disk gas to diffuse through $\Delta a_{\rm hr}$ is much smaller than 
$\tau_{\rm lib}$ such that a large fraction of the gas diffuses through 
their horseshoe regions without being perturbed by their gravity. 
In the high-mass limit, the vortensity and entropy related part 
of corotation torque is saturated. In the low-mass limit, 
 the magnitude of corotation torque approaches the 
linear corotation torque which is below those of differential Lindblad torque. 
The embryos would undergo inward migration throughout the disk.

Within certain mass range ($\sim$ a few $M_\oplus$), embryos undergo convergent 
type I migration towards transitional radii ($r_{\rm trans}$) where the 
net (Lindblad plus corotation) torque changes sign with a negative radial 
gradient. It has been suggested this process may lead to the accumulation 
of building block material and enhance the growth of embryos \citep{
2005A&A...443.1067N, 2010ApJ...715L..68L, 2011MNRAS.410..293P, 2012ApJ...750...34H, 
2012MNRAS.419.2737H}. 
\cite{2013A&A...558A.105P}  investigated the effects of initial number of embryos 
and the stochastic force on the frequency of Gas Giants’ formation.
 They suggested that the resonant chain can be broken by increasing the initial number 
of embryos or by including a moderate stochastic force due to the disk turbulence. 
They also discussed briefly about the dependency of zero-torque radius on the 
protoplanetary disk mass, but the effects of different disk mass on convergent migration 
of embryos were not discussed. 
In this paper, we  emphasize on this potential mechanism and explore conditions 
under which the critical mass cores ($\sim 10 M_{\oplus}$) may form and be retained 
during the main course of disk evolution.
After we submitted our paper, a paper by \citet{2014arXiv1408.6993C} was
posted on Arxiv.org. They  investigated the correlation between initial 
disk mass and the formation and survival of gas giants, and  pointed out 
that in order to form and retain gas giants, it is necessary for planetary 
cores to accrete gas and open gap at large radii and they must do it in a 
sufficiently late epoch to prevent migrating into the central star. However, 
during this advanced stage, much of the initial disk mass is depleted from 
its initial values so that the critical condition for cores’ retention by 
their natal disks is not yet explicitly determined. Although we obtain some 
similar results on the condition for multiple embryos to overcome resonant 
barriers (as found by \citet{2014arXiv1408.6993C}), our consideration of 
more general boundary conditions in this and subsequent papers bypass 
some retention issues associated with their corotation saturation. 
Following previous work, we use in the paper a 2D FARGO code 
to carry out hydrodynamic simulation of tidal interaction between multiple 
embryos and their natal disks. In \S2, 
we briefly recapitulate the numerical method and model parameters.

Following the detailed analysis by \citet{2010MNRAS.401.1950P, 
2011MNRAS.410..293P}, we verify in \S\ref{sec:single} that in 
disks with {\it composite} $\Sigma_g$ distribution, isolated 
embryos undergo the type I migration to $r_{\rm trans}$. These 
simulations are extended to multiple embryos in \S\ref{sec:trace}.  
Since contribution from corotation torque determines the 
direction of type I migration, we introduce an idealized model to
examine whether it may be affected by overlapping horseshoe regions 
between two nearby embryos. In this model, we neglect mutual gravitational
interaction between embryos and trace diffusion across the corotation
zone with passive contaminants.  

As they approach each other, embryos perturb each other through
secular and resonant interaction.  These effects are restored in
the models presented in \S\ref{sec:multi}.  We show that in disks with modest
$\Sigma_g$, convergent embryos are trapped into their mutual mean 
motion resonance.  But embryos' rate of type I migration increases
with $\Sigma_g$ and in relatively massive disks they bypass the 
resonant barrier.

After their orbits cross, embryos undergo close encounters.  
We show in \S\ref{sec:multi} that although some embryos scatter each other 
to location outside the corotation zone, they resume their 
convergent type I migration. These embryos are entrenched 
near $r_{\rm trans}$ by persistent type I migration and undergo 
repeated close encounters until they collide with each other.
In the 2D simulations, physical collisions occur within a 
few hundred orbital periods.   However, if the embryos have an
nearly isotropic velocity dispersion rather than mono layer distribution, 
their collision time scale is 2-3 order of magnitude longer, ie a 
significant fraction of the (Myr) disk evolution time scale. It is
impractical to carry out high-resolution hydrodynamic simulation 
of embryos' close interaction during various stages of 
disk evolution. In a subsequent paper, we will utilize a 
prescription for embryo-disk interaction to construct a Hermit-Embryo 
scheme.  This prescription was constructed by
 \citet{2010MNRAS.401.1950P, 2011MNRAS.410..293P} 
based on the results of a comprehensive series of hydrodynamic 
simulations of the embryos-disk interaction. This approximation 
is justified under the assumption that embryos do not strongly modify 
the disk's intrinsic structure. Based on the results here, we assume
this prescription remains valid for individual embryos in multiple 
systems. Finally, in  \S\ref{sec:summary}, we summarize our results and 
discuss their implications.


\section{Hydrodynamic Simulation of Embryos-Disk Interaction}
\label{sec:method}


Here, we briefly
recapitulate the numerical method and model parameters.


Following previous investigations, we utilize a publically available FARGO 
(Fast Advection in Rotating Gaseous Objects; \citet{2000A&AS..141..165M}) 
scheme to simulate the interaction between multiple embryos with their 
natal disks.  FARGO is a 2D hydrodynamical polar grid code centred
on the star, based on the van Leer upwind algorithm on a staggered mesh.
It solves the Navier-Stokes and continuity equations for a Keplerian disk 
subject to the gravity of the central object and that of embedded 
protoplanets as well as the energy equation in its more recent version 
\citep{2008IAUS..249..397B}. The energy equation implemented in FARGO is:
\begin{equation}
\frac{\partial e}{\partial t}+\vec{\nabla}\cdot(e\vec{v})=-P\vec{\nabla}\cdot\vec{v}+Q_{+}- Q_-
\end{equation}
where $e$ is the thermal energy density (thermal energy per unit area), 
$\vec{v}$ is the flow velocity, $P$ is the vertically integrated pressure, 
and $Q_+$($Q_-$) denote heating (cooling) source terms, assumed to be 
positive quantities. The cooling source is defined by a cooling time, 
the disk will go back to the initial energy within the given cooling 
time. Here we choose the cooling time to be about 5 orbital periods of 
the planet at $r=1$.

There're several boundary conditions available in the FARGO code. Here 
we choose the EVANESCENT boundary condition which is described in 
\citet{2006MNRAS.370..529D}, used for the EU test comparison problem. 
It aims at implementing wave killing zones at each edge of the grid. 
It allows the disk values (surface density, velocities, thermal energy density) 
to damp toward the instantaneous axisymmetric disk conditions. 
The damping regions are located in the radial ranges [$r_{min},1.25r_{min}$] 
and [$0.84r_{max},r_{max}$], where $r_{min}$($r_{max}$) denotes the inner 
(outer) edge radius of the grid.

The basic equations are solved in a cylindrical coordinates from $r=0.3$ 
to $r=1.7$ and full $2\pi$ in azimuth. The typical resolution is 
$\delta r/r_p\sim 0.004$ and $\delta\phi=0.01$, which gives about 5 grids 
in radial direction within the Hill's radius of a $10M_\oplus$  planet.
\vspace{3ex}

For illustration, we adopt a simple $\alpha$ disk model with an 
effective viscosity.  We assume power-law distributions
\begin{equation}
\frac{\partial ln\Sigma_g}{\partial ln{r}}=p, \ \ \ \ 
\frac{\partial ln{T}}{\partial ln{r}}=q, \ \ \ \
\frac{\partial ln\alpha}{\partial ln{r}}=\zeta.
\end{equation}
In a steady state, the accretion rate $\dot{M}=3{\pi}\Sigma_g
\alpha{C_s}^2\Omega$ is independent of $r$ so that $p+q+\zeta
=-1.5$.

We adopt a disk model based on the assumption that 
the inner region of the disk is heated by viscous dissipation 
whereas the outer region is heated by stellar luminosity \citep{2007ApJ...654..606G}.
we set the transit region at $r_t$, the 
disk inside $r_t$ has $p=0.0$ and $q=-1.5$ 
 and outside has $p=-0.5$ 
and $q=-1.0$. In the simulations, these model parameters follow a
 continuous transition at $r_t$.  This composite power-law $\Sigma_g$ 
and $T$ distribution model is an extension of the single power-law 
simulations pioneered by \citet{2010MNRAS.401.1950P, 2011MNRAS.410..293P}. 

We simulate the planetary migration in disks of both high and low accretion rate 
to investigate the dependency of orbital structure on the convergent migration rate.
According to the disk model structured by \citet{2007ApJ...654..606G}, 
$h/r \propto \dot M^{0.25}$ and $r_t \propto \dot M^{0.72}$. 
For normalization, we specify 
$h/r=0.079$ at $r_t$ for a disk with accretion rate as high as $10^{-7}M_{\odot}yr^{-1}$,
 and $h/r=0.05$ at $r_t$ for $1.2 \times 10^{-8}M_{\odot}yr^{-1}$ (model 5). 
To reduce the simulation time for planetary migration in low $\dot M$ disks, 
we set $r_t=7.1$ AU for all disk models. 
The viscosity $\alpha$ is $0.001$ as constant, and the smoothing length $b/h=0.4$.

We first simulate the migration of a single isolated planet in a disk of 
composite $\Sigma_g$ or $S_g$ distributions to confirm that 
the planets would undergo convergent migration in this kind of disk model (model 1a-1f). 
Then we compare the results with the migration of a single planet in a disk disturbed by other
planets (model 1g). 
At last, we simulate the migration of multiple planet systems in disks with different
accretion rates (model 5 \& 6). 
In all models, the planet has initial circular orbit.  The model parameters are listed in Table 1.

\begin{deluxetable}{cccccccc}
   \tablecolumns{6}
   \tablehead{\colhead{model} & \colhead{$N_p$} & \colhead{$M_p (M_\oplus)$} & \colhead{$\dot M (M_\odot  yr^{-1})$} & \colhead{feel others} & \colhead{feel disk}}
   \startdata
         1a-1c & 1 & 10  & 1e-7 &  NO  & YES\\
         1d & 1 & 20  & 1e-7 &  NO  & YES\\
         1e & 1 & 30  & 1e-7 &  NO  & YES\\
         1f & 1 & 40  & 1e-7 &  NO  & YES\\
         1g & 3 & 10  & 1e-7 &  NO  & YES\\
         5 & 4 & 10  & 1.2e-8 &  YES  & YES\\
         6a & 4 & 10  & 1e-7 &  YES  & YES\\
         6b & 4 & 10  & 2e-7 &  YES  & YES
    \enddata
    \tablenotetext{}{feel other: gravitational interaction with other planets.}
    \tablenotetext{}{feel disk: interaction between planet and disk.}
    \label{tab:simulation}
\end{deluxetable}

\section{Migration of single isolated planets}
\label{sec:single}

\begin{figure*}[!tbp]
 \centering
\includegraphics[scale=0.4, angle=0]{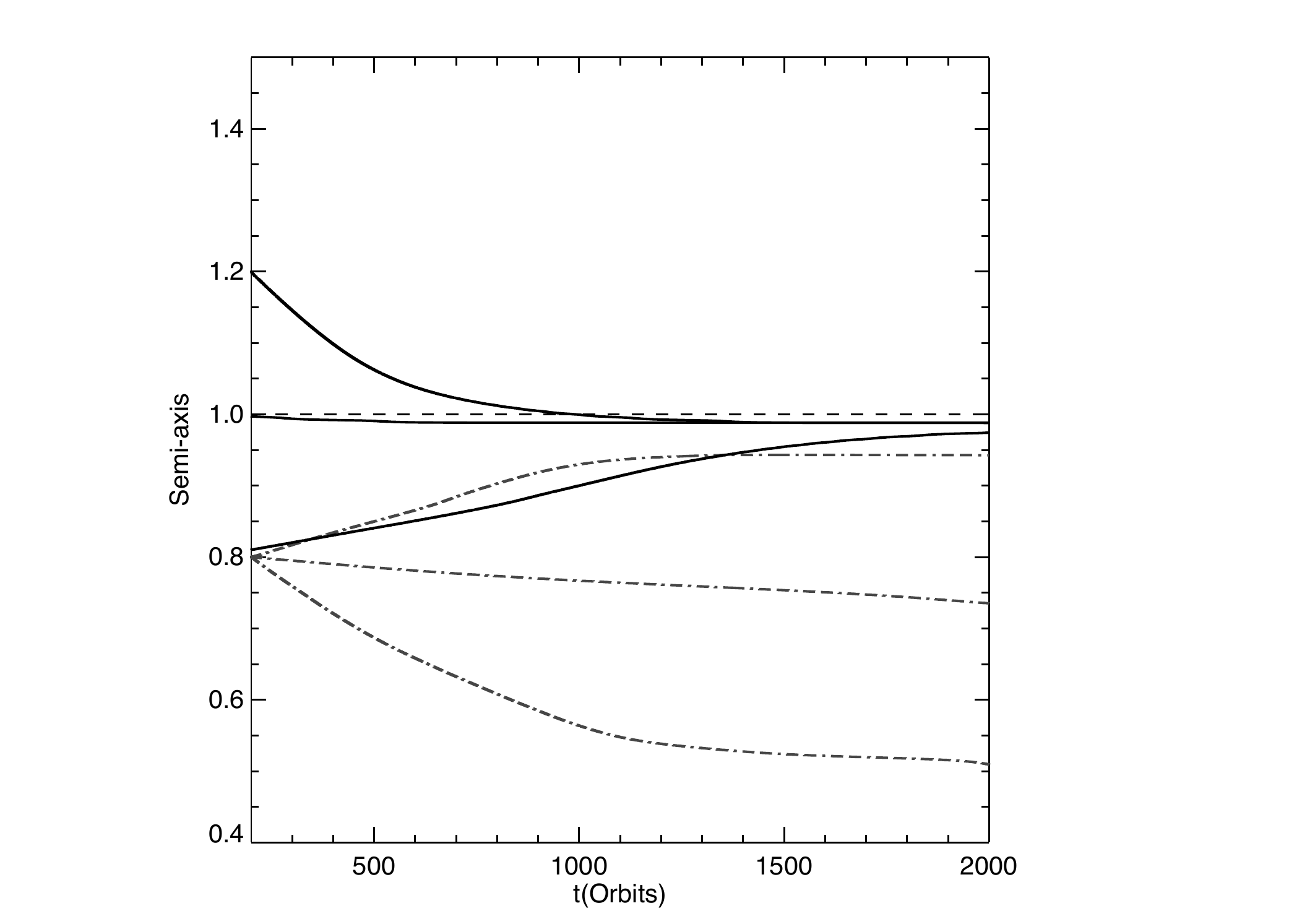}
\includegraphics[scale=0.4, angle=0]{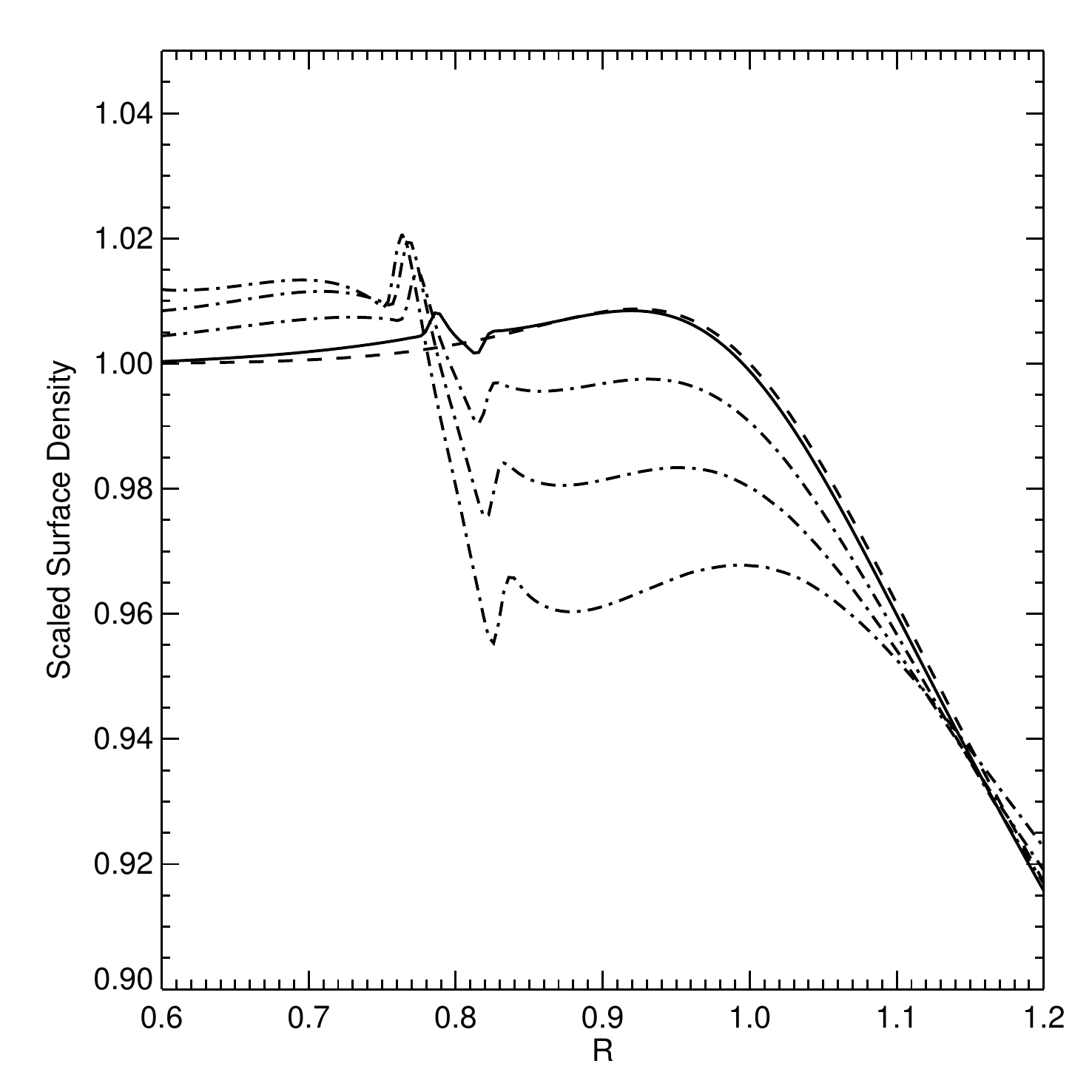}
\caption{\textbf{Left panel}: The three solid lines indicate the
semi-axis evolution of a $10M_\oplus$  
planet released from $r=0.8$, $r=1.0$ and $r=1.2$ respectively(model 1a-1c). 
The three dot-dashed lines indicate the
semi-axis evolution of a planet from $r=0.8$ with mass of $20$, 
$30$ and $40M_\oplus$ respectively(model 1d-1f). 
 The dashed line indicates the location of $r_t$. 
\textbf{Right panel}: Surface density profile scaled with unperturbed $\Sigma(r_t)$. 
The dashed line indicates the profile of unperturbed disk. The solid and three dot-dashed curves,
 with value at $r_t$ from higher to lower,  indicate that disturbed by a planet with mass of 
$10$, $20$, $30$ and $40M_\oplus$ respectively(model 1a, 1d-1f) after 200 orbits.
}
\label{fig:2s1p}
\end{figure*}

The planet with 
$M_p = 10M_\oplus$ migrate outward when released inside 
the transit location (model 1a) and migrate inward when 
starting outside (model 1c) (Fig.~\ref{fig:2s1p}). The sign and 
magnitude of the initial torque agree well with that obtained 
for a planet in a disk with a single power-law $\Sigma$ 
distribution. In both cases, the migration is slowed down 
and stalled when the planet approaches $r_t$. Although
the strength of one-side torque is preserved, the net Lindblad 
torque is suppressed by the cancellation from the two regions of 
the disk. The reduction of the differential Lindblad torque 
enables it to balance the corotation torque. These models indicate 
that in disks with a bimodel ($p$, $q$) distribution, the net 
torque can indeed be summarized as a linear combination of that
from two separate regions of the disk.

In model 1d-1f, we place a single isolated planet at initial location $r=0.8$
with $M_p=20 M_\oplus$, $30 M_\oplus$ and $40 M_\oplus$ respectively.  
The planet migrated outward with stalling at $r_{t}$ in model 1d 
and migrated inward in model 1e and 1f (Fig. \ref{fig:2s1p}).
 For this $M_p$, perturbation on $\Sigma_g$ remains
relatively small. These results are in general agreement with 
those obtained by \citet{2010MNRAS.401.1950P, 2011MNRAS.410..293P} and 
they indicate that the corotation torque of massive planets is 
saturated (weakened) by the suppression of diffusion of disk gas 
across the horseshoe stream lines.


\section{Migration of multiple planets}
\label{sec:trace}

When two or more planets converge near $r_t$, the location of their 
Lindblad resonances and horseshoe region may overlap with each other.
They also directly interact with each other.  We assess the relative
contribution from different effects with idealized simulations of 
the concurrent evolution of multiple planets.

\begin{figure*}[!tbp]
 \centering
\includegraphics[scale=0.4, angle=0]{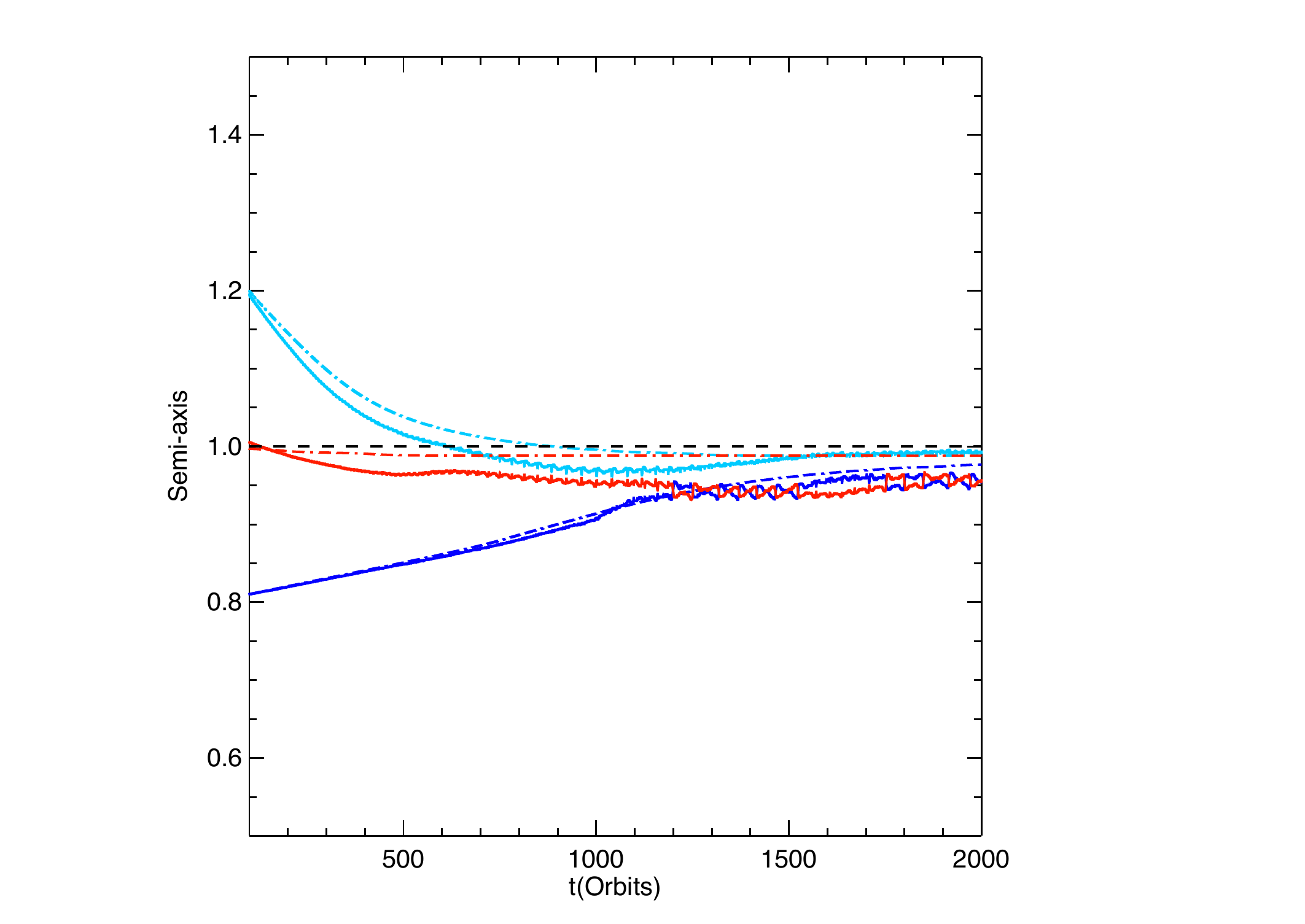}
\includegraphics[scale=0.395, angle=0]{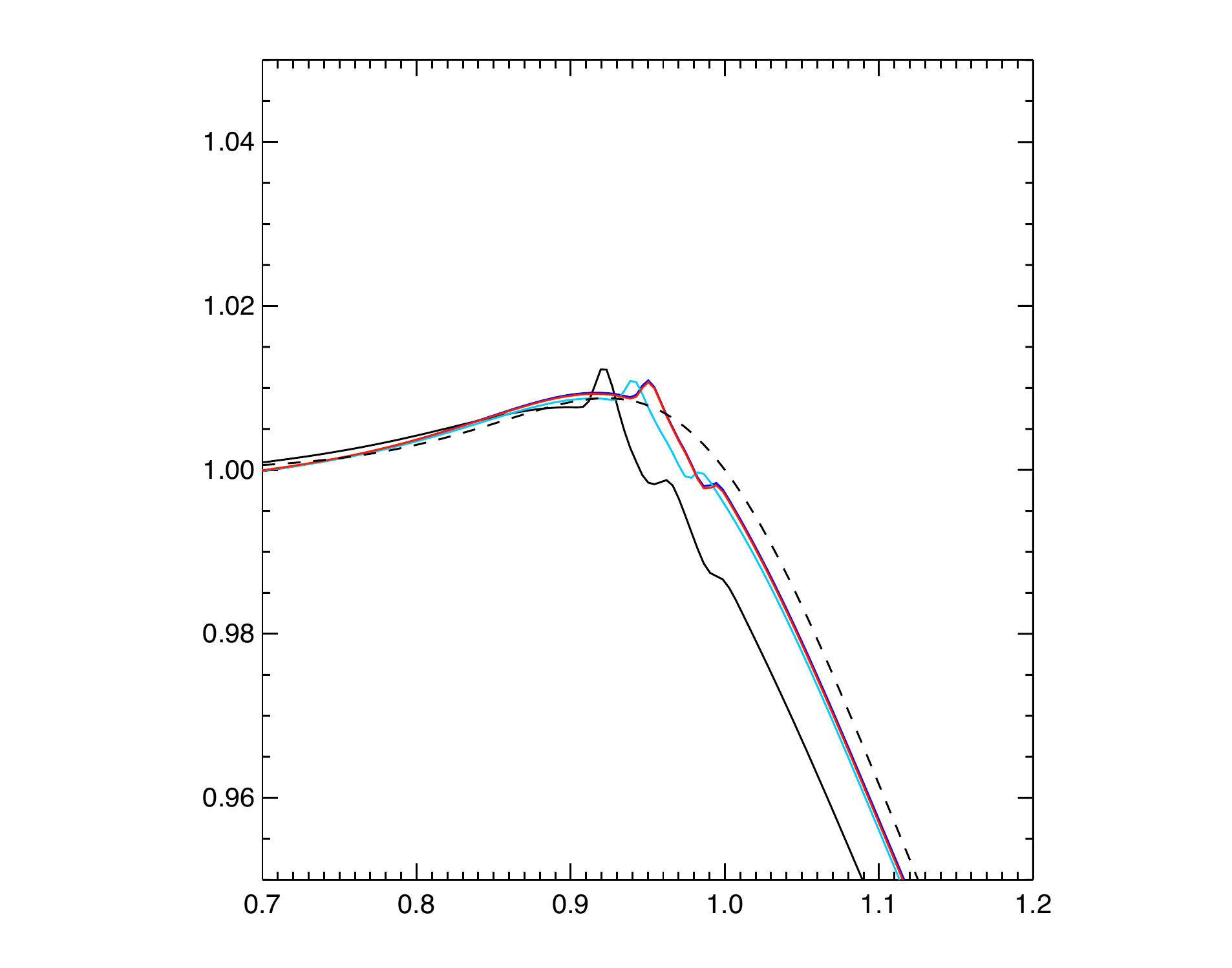}
\caption{\textbf{Left panel:}
The semi-axis evolution of three planets with equal mass of $10M_\oplus$ (model 1g).  
The planets are released from $r=0.8$ (dark blue solid), 
$r=1.0$ (red solid) and $r=1.2$ (light blue solid). 
During their migration, planets' direct gravitational perturbation on
each other is neglected. However, the disk is perturbed by all three 
planets. The dot-dashed color lines represent the migration with single 
planets from previous comparison models 1a-1c. The dashed black line indicates 
the location of $r_t$.
\textbf{Right panel:} The dashed solid and black lines represent the 
surface density profile at $t=0$ and $t=2000$ orbits in the case of 
multi-planets' migration respectively. The color lines represent 
the surface density profile at $t=2000$ orbits of the previous three 
separate cases of isolated planet's migration. 
}
\label{fig:2s1p3p}
\end{figure*}

We first examine the extent of resonant interference by releasing
three planets at the same location as the previous single-planet 
models. In an artificial, idealized model, we neglect the mutual 
interaction between the planets (model 1g).  Initially, the disk response is 
a linear combination perturbation of three widely separated planets.
These planets independently evolve along paths similar to those 
of the analogous individual planets (Fig.~\ref{fig:2s1p3p}). As they
approach $r_t$, their combined perturbation amplitude on the 
disk is locally enhanced, albeit changes in $\Sigma_g$ 
remains small and linear (Fig.~\ref{fig:2s1p3p}).

 
\begin{figure*}[htbp]
\includegraphics[scale=0.82, angle=0]{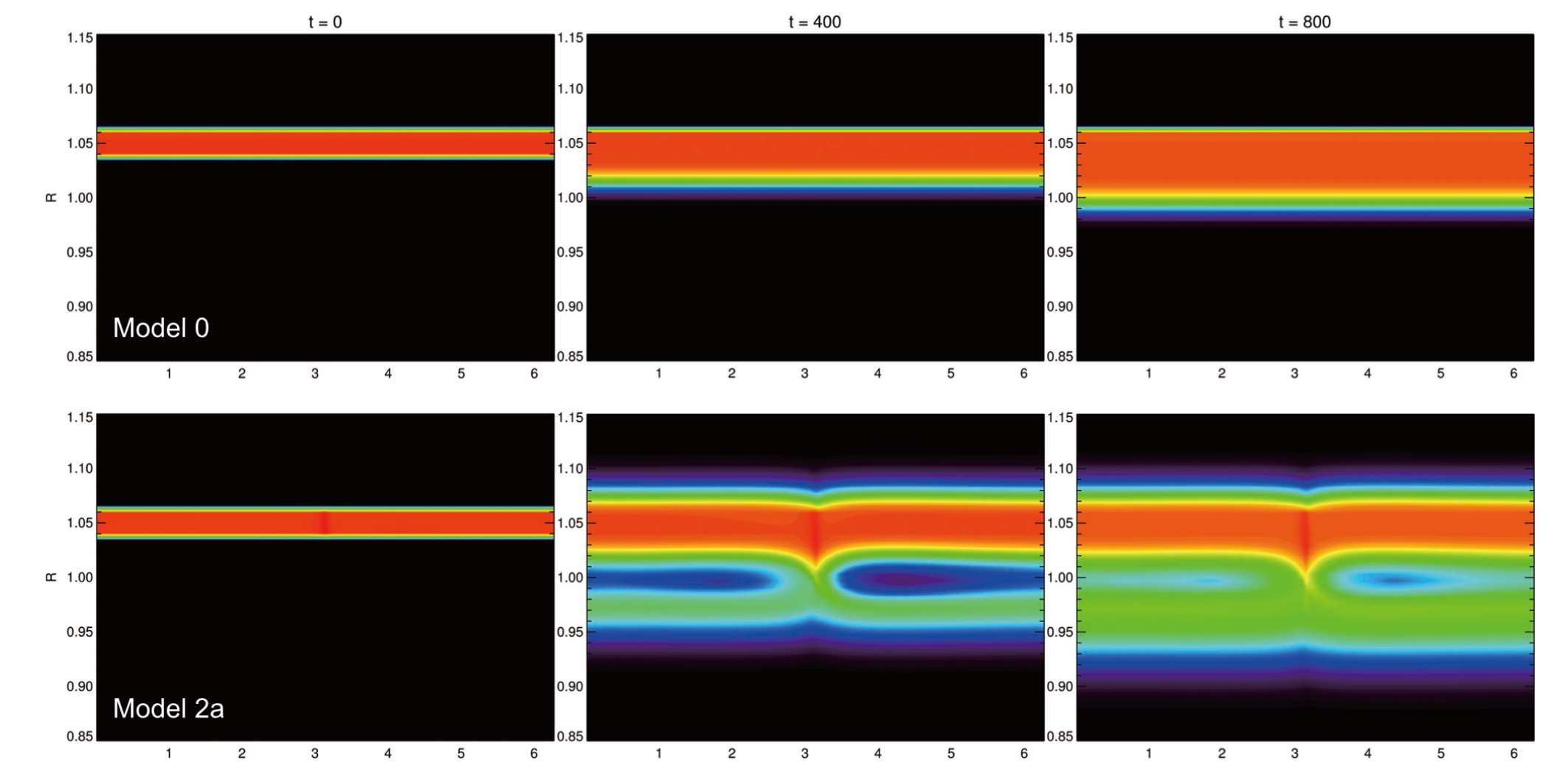}
\includegraphics[scale=0.82, angle=0]{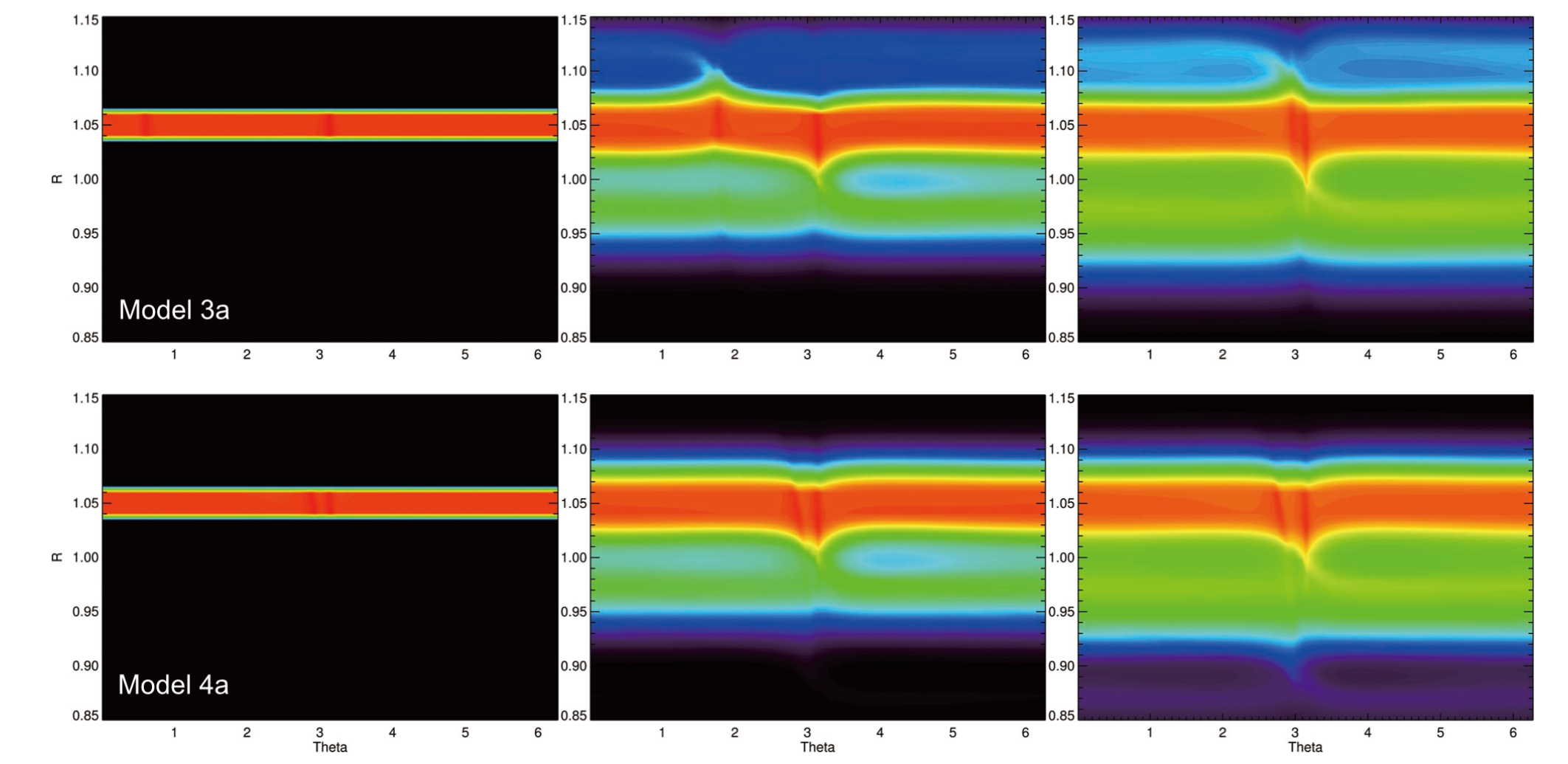}
\caption{
Diffusion of tracer elements in an unperturbed disk  model 0
({\bf row 1}), a disk with one planet fixed at $r_t$ in model 2a 
({\bf row 2}), a disk with two planets fixed outside $r_t$ in model 3a
({\bf row 3}), and two planets fixed inside $r_t$ in model 4a 
({\bf row 4}). All the planet has equal mass as $10M_{\oplus}$.  
The left column represents the initial tracer distribution.  The 
middle and right columns represent the tracer
distribution at t=400 and 800 orbits.}
\label{fig:tracer1}
\end{figure*}

In order to further investigate the interference of planets 
on each other's migration, we introduce a tracer (as passive
contaminant) to highlight the diffusion of fluid elements 
around the planets when they are very close to each other. 
For comparison, the fluid elements' tracer is examined 
under four situations: (model 0) an unperturbed disk 
(without any embedded planet), (model 2a) a disk which bears 
a single planet with a fixed position at $R=1.0$, (model 3a) 
disk with two planets fixed at $R=1.0$ and $R=1.1$ and 
(model 4a) disk with two planet at $R=1.0$ and $R=0.9$.
Figure~\ref{fig:tracer1} shows the diffusion pattern of tracer 
elements under these 4 cases with $M_p=10M_{\oplus}$.  Models 
1a-4a are presented from top to the bottom rows.  The 
tracer distribution at time 0, 400, and 800 orbital period 
are plotted in left to right columns.  

Comparisons between models 0 and 2a highlight the 
horse shoe stream lines near the corotation region around an 
embedded planet.  Width of the corotation zone around the 
$10 M_\oplus$ planet is $\sim 0.05$.  In models 3a and 4a,
the corotation zones of the two planets are adjacent to
each other without any significant overlap. Nevertheless,
their co-existence leads to weak diffusion to either 
outer (model 3a) or inner (model 4a) regions.

\begin{figure*}[htbp]
\includegraphics[scale=0.82, angle=0]{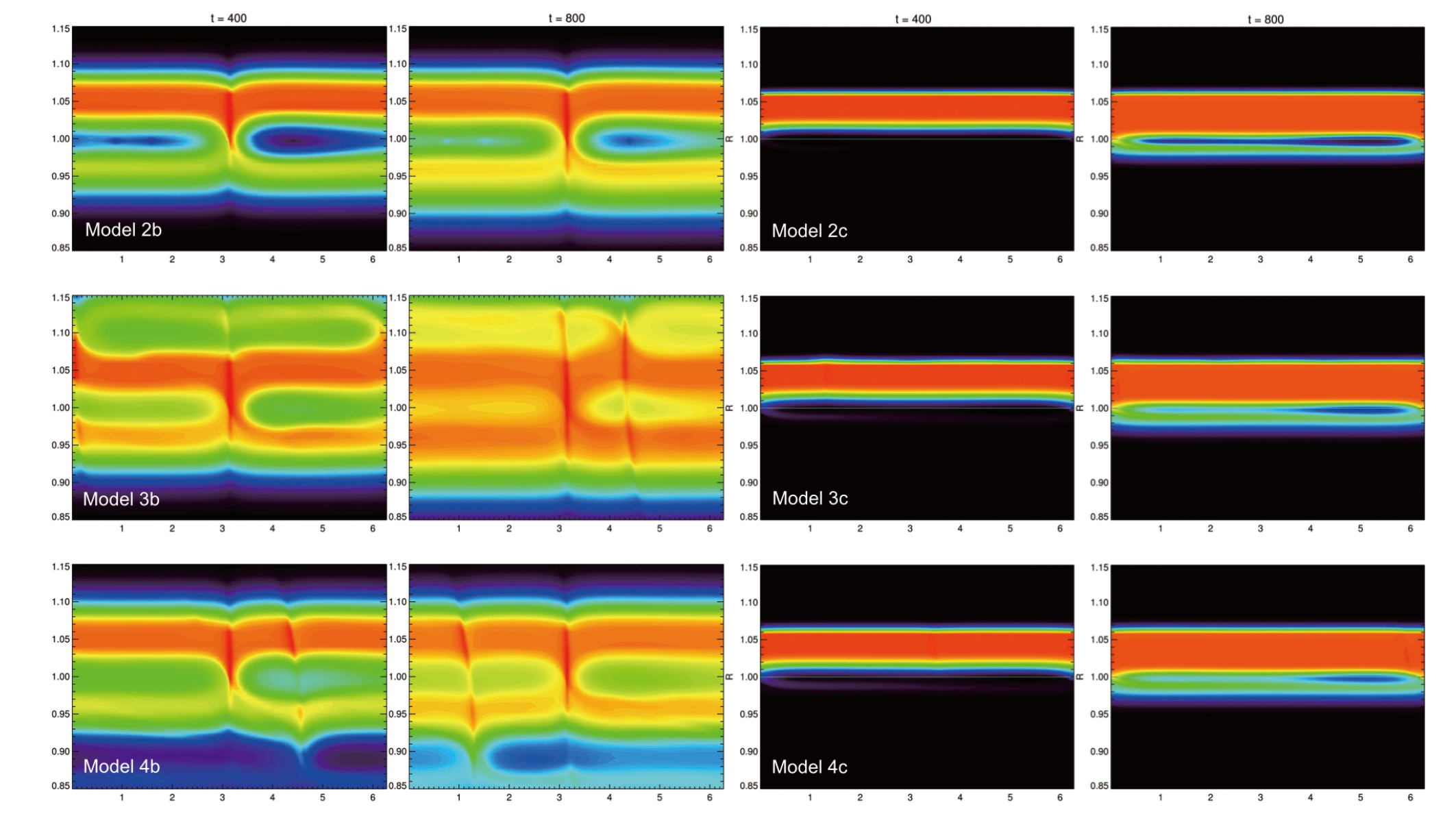}
\caption{
  Diffusion of tracer elements for models 2b-4b ({\bf rows 1-3}) with 
20 $M_\oplus$ planets after 400 and 800 orbits in columns 1 and 2 from
the left.  Similar disk models 2c-4c with 2 $M_\oplus$ planets after
400 and 800 orbits are shown in columns 3 and 4. 
}
\label{fig:tracer2}
\end{figure*}

Figure~\ref{fig:tracer2} shows the tracer distribution for models (2b-4b) 
with $M_p=20M_{\oplus}$ (in rows 1-3).  Columns 1 and 2 (from
the left) represent the distribution after 400 and 800 orbits.
Similar models (2c-4c) with $M_p=2M_\oplus$ at the corresponding epochs
are plotted in columns 3 and 4 (from the left).  

The corotation region around a $2 M_\oplus$ planet (model 2c) 
is $\sim 0.02$. The corotation regions around the coexisting 
planets in models 3c and 4c are well separated.  There is no 
evidence of horse shoe stream line interference to enhance 
the diffusion of tracer elements.

In contrast, model 2b shows that around a 20$M_\oplus$ planet, the 
width of the corotation region is $\sim 0.07$ such that two planets 
with 0.1 separation have overlapping corotation regions. The horse 
shoe stream lines around two planets are clearly intertwine in 
models 3b and 4b.  Gravitational perturbation from the exterior 
planet enhances outward diffusion whereas that from the interior 
planet promotes more rapid inward migration.  

In previous analysis, \citet{2010MNRAS.401.1950P, 2011MNRAS.410..293P} 
showed that
the preservation of the $\Sigma_g$ profile near the embedded planets
and the maintenance of their corotation torque require gas diffusion 
through the corotation region. For relatively large planets, the 
diffusion is quenched by the horse shoe streamlines and saturates the
corotation torque. The elevated diffusion of tracer elements in disks 
with closely packed multiple planets (models 2c and 2d) limits the
modification of the unperturbed $\Sigma_g$ distribution and unsaturates 
the corotation torque in comparison with that between the disk and 
isolated planets (model 2b).  

Our simulated models suggest that 1) the migration direction, pace
and the net torque of individual planets are not affected by other
planets in well separated multiple systems and 2) closely-packed 
multiple planets also retain their unsaturated corotation torque.
Thus, resonant interference enhances rather than suppresses convergent
migration.  Based on these results, we approximate, in subsequent papers 
of this series, the torque exerted by individual planets in multiple 
systems with the prescription obtained from simulations with single 
planets \citep{2010MNRAS.401.1950P, 2011MNRAS.410..293P}.

\section{Gravitational interaction and merger through physical collisions}
\label{sec:multi}

We now incorporate the gravitational interaction between planets. 
We place four equal-mass ($10M_\oplus$) planets into the disk (with the same
$p, q$ distribution as previous models) at $r=0.7, 0.9, 1.1, 1.3$ 
respectively.  

\begin{figure}
 \centering
\subfloat{\includegraphics[width=0.75\linewidth,clip=true]{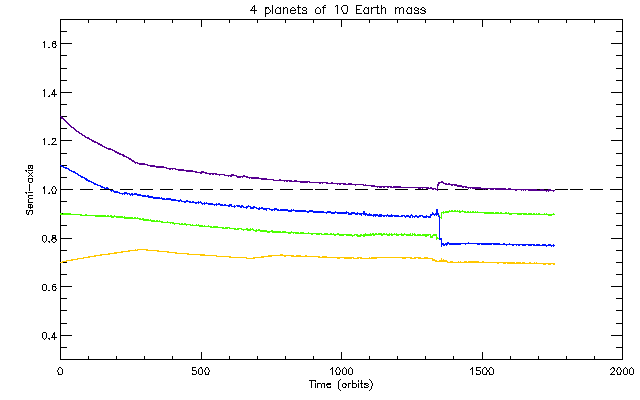}} \\
\subfloat{\includegraphics[width=0.75\linewidth,clip=true]{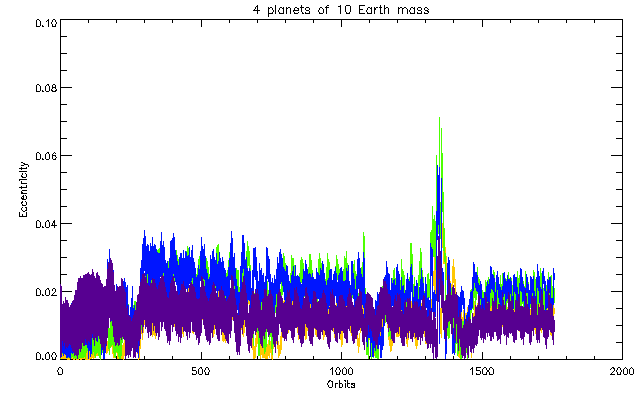}} 
\caption{
The semi-axis (\textbf{top}) and eccentricity (\textbf{bottom}) evolution of four $10M_\oplus$ planets' orbits 
released from $r=0.7, 0.9, 1.1, 1.3$ respectively in a disk of  $\dot M=1.2 \times 10^{-8} M_{\odot}/yr$ (model 5) .
}
\label{fig:4pAElowM}
\end{figure}

In model 5, we specify a low-mass disk with $\dot M=1.2\times10^{-8} 
M_{\odot}yr^{-1}$. In this case, the embryos rapidly capture each other on 
their 6:5 or 5:4 mean motion resonances (Fig.~\ref{fig:4pAElowM}).  This 
resonant configuration is maintained with a fluctuating eccentricity 
$e \sim R_R/a \sim 0.02$ while the convoy's inward migration slows down 
and comes to a halt.  In this compact configuration, the periastron of some 
embryos crosses the apoastron of other embryos which are slightly closer 
to their host stars.  After 1300 orbits, two middle embryos' orbit 
overlapped and they exchange their orbit to re-establish the resonant 
configuration.  In general the embryos preserve their integrity and may
undergo further migration together as the disk is depleted.  We suggest
that these convoys are the progenitors of multiple super Earths.

We also introduce models 6a and 6b with similar embryos' mass and 
initial location as model 5.  But the gas accretion in the disk is 
specified to be $10^{-7}$ and $2 \times 10^{-7} M_\odot yr^{-1}$
respectively.

\begin{figure}[!tbp]
 \centering
\subfloat{\includegraphics[width=1.2\linewidth,clip=true]{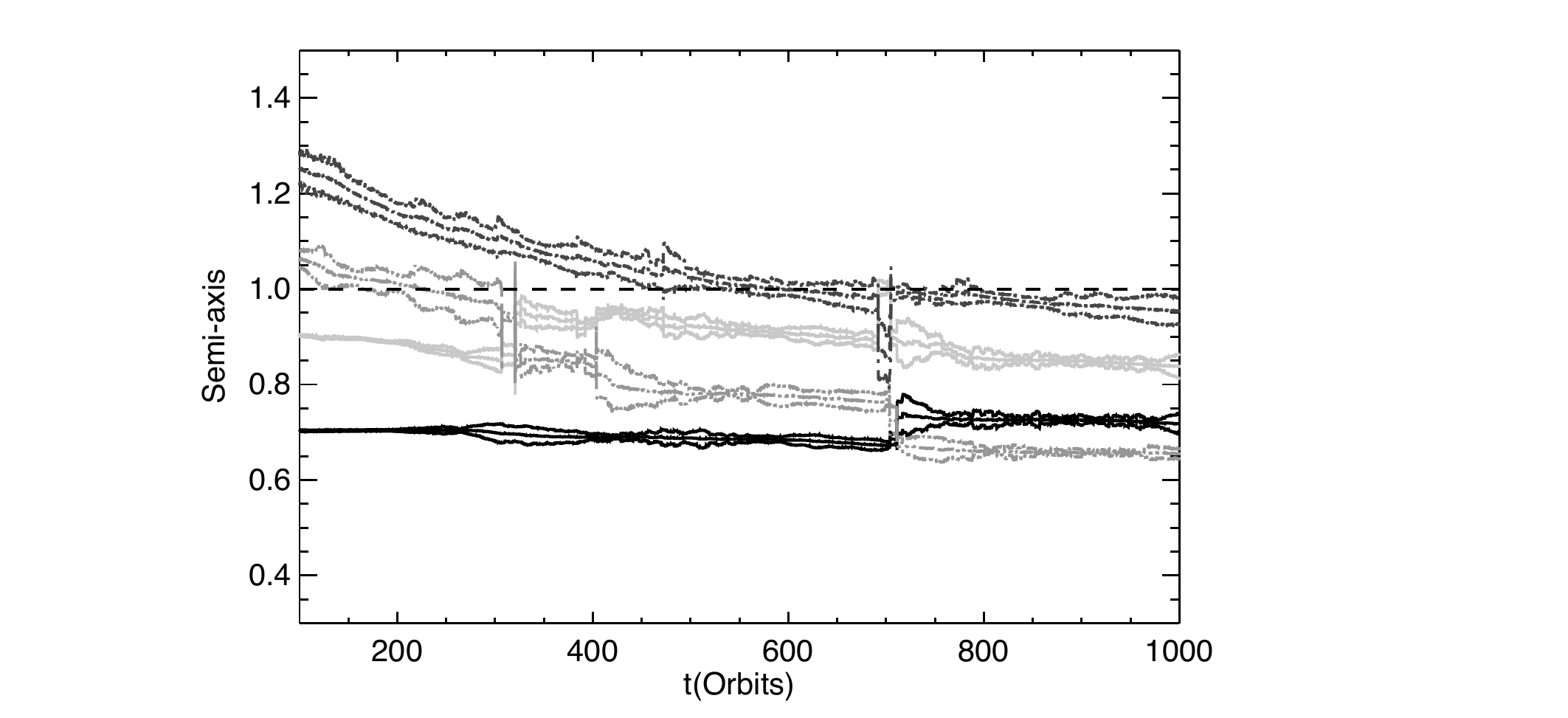}} \\
\subfloat{\includegraphics[width=1.2\linewidth,clip=true]{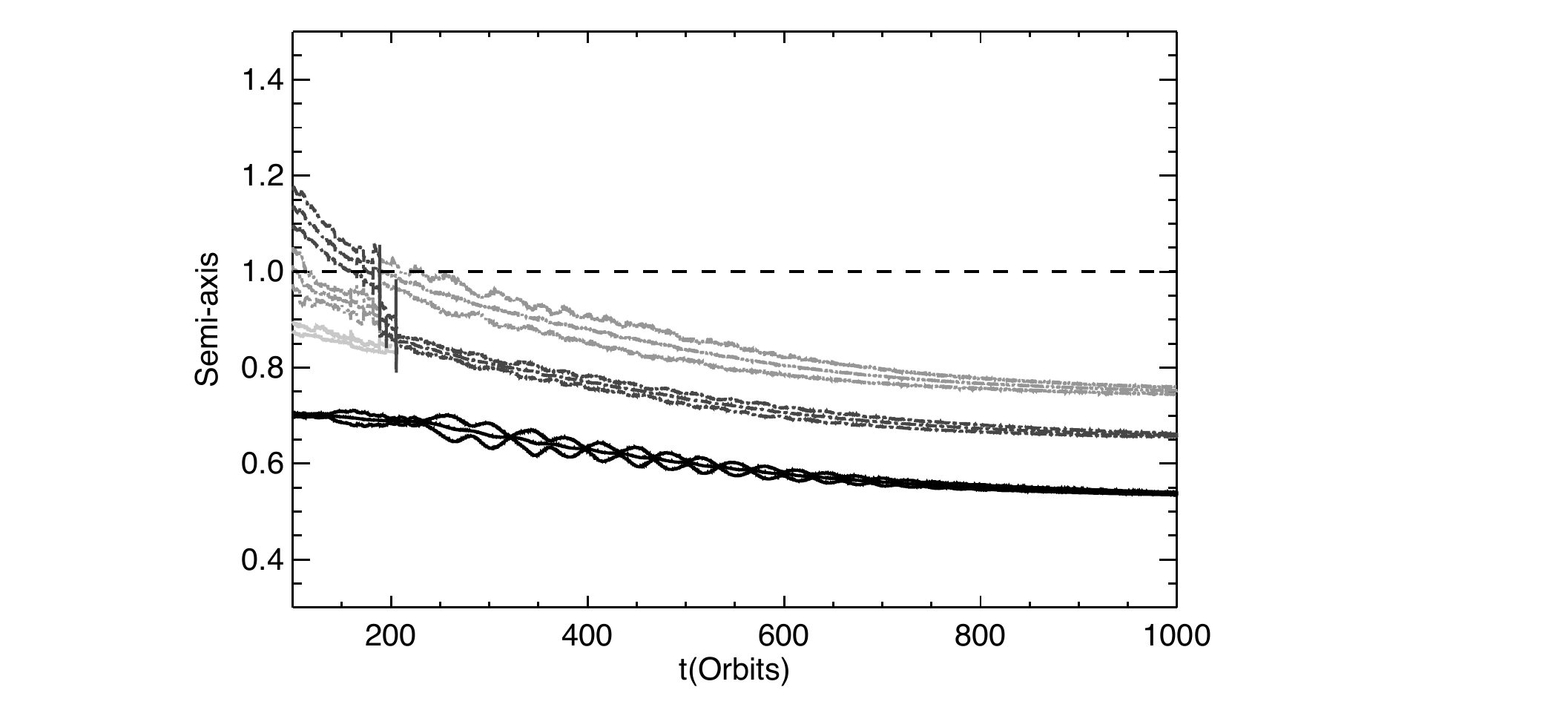}} 
\caption{
The semi-axis and eccentricity evolution of four $10M_\oplus$ planets' orbits  
released from $r=0.7, 0.9, 1.1, 1.3$ respectively in disks with different accretion rate. 
Each group of curves show the paths of $a$, $a \times (1+e)$ and $a \times (1-e)$ for the same planet.
(\textbf{Top}: $\dot M=10^{-7} M_{\odot}/yr$(model 6a), \textbf{Bottom}: $\dot M=2 \times 10^{-7} M_{\odot}/yr$(model 6b)).
}
\label{fig:4pAE}
\end{figure}

Similar to the results of model 5 and our previous simulation \citep{2014ApJ...789L..23Z},
 embryos in model 6a first capture each other 
into lower-order mean motion resonances.  During their subsequent 
collective migration, their separation is reduced to about half of their initial spacing. 
Embryos' $e$ fluctuates mostly with an 
average amplitude $<0.02$ as it is excited by the embryos' resonant 
interaction with each other and damped by the tidal torque between them 
and the gas. 
Since their separation is $\sim 0.1 r_p$, their orbits 
do not generally cross each other.  But, on some occasions (eg at $\sim 300$
orbits), pairs of closest embryos may become dynamically unstable to
undergo orbit crossing.  Close encounters lead to the exchange of semi
major axes and eccentricity excitation ($\sim R_R/a$ up to $\sim 0.05$) 
(Fig.~\ref{fig:4pAE}).  After this brief episode of close encounters
and intense interaction, embryos'
eccentricity is quickly damped by the tidally induced gas drag 
from the disk gas. 

A convoy of embryos settles near but not precisely 
around $r_t$.  This slight asymmetry is due to the embryos' torque 
balance between the inner and outer regions of the disk.  After 
$\sim 700$ orbits, the two inner most embryos again cross each 
other's orbit and exchange their semi major axes.  
Models 5\&6a indicate the possibility of repeated orbit crossing.  
Since embryos would undergo orbit crossing when their separation is 
smaller than the width of feeding zone \citep{2007ApJ...666..423Z}, 
under some circumstances, the embedded embryos may 
converge into regions where they undergo repeated 
close encounters. Some encounters may be sufficiently
close that the participating planets physically
collide and merge with each other. In this series of
papers, we will explore the possibility that these
merger may attain critical mass for the onset of 
efficient gas accretion.  In order to simulate this possibility 
with FARGO,  we remove the gravitational softening parameter 
in the calculation of the force between the participating embryos
and assume they would merge
with each other, with the conservation of mass 
and momentum, if their impact parameter is smaller
than a critical merger size $R_c \simeq 4 R_p$. 

In a previous analysis \citep{2014ApJ...789L..23Z}, we showed that the embryos' 
convergent type I migration is halted with an interplanetary spacing 
$\Delta a$ which is a decreasing function of ${\dot M}$.  In a 
steady disk with a constant $\alpha$, $\Sigma_g$ generally increases
with the accretion rate ${\dot M}$.  The strength of both differential
Lindblad and corotation torque as well as the embryos' type I migration 
speed increase with $\Sigma_g$.  We introduce model 6b to show that in
disks with relatively large $\Sigma_g$ (or equivalently ${\dot M} = 2 
\times 10^{-7} M_\odot$ yr$^{-1}$), the embryos
undergo the similar paths as that in the previous models. 

We find that, the faster convergent migration rate leads to the merge of the middle 
two embryos shortly after released. Later they converged into a resonant state 
(Fig.~\ref{fig:4pAE}). 
If we use the embryos' actual physical size $R_p$ with
a comparable density as the Earth, the estimated
collision time scale would be $\tau_c \sim 
(a \delta a / N R_p^2 \Theta) P$, which is several orders 
of magnitude longer than their orbital period $P$ 
even after they converge into a region with a radial 
width $\delta a$ comparable to their Roche radius 
or the extent of their corotation region. 
However, under the 2-D disk simulations, the collision frequency is much higher, 
the orbital crossing time scale could be reduced to several hundred orbital period. 
Realistic collisions require protracted simulations which is
beyond the current computational constraint.

The total mass of the embryos (40 $M_\oplus$) in models 5, 6a and 6b
exceeds $M_p$ ($=20 M_\oplus$) in models 1d-1f (see \S\ref{sec:single}). 
Yet, this convoy of resonant embryos are retained near $r_t$ whereas
the more massive isolated embryo migrated toward the inner boundary
of the disk.  These results indicate that, as separate entities, 
lower-mass embryos preserve their corotation torque 
(in model 6a).  But if they merge into sufficiently massive isolated 
embryos, their corotation torque would be saturated and they may
no longer be trapped near $r_t$ (model 6b). 
After the merger event, the massive embryo continues to migrate inward. 
Its perturbation on the disk structure is enhanced.
In isolation, the merger's mass is sufficiently large for it 
to confine the horse shoe stream lines and saturate the 
corotation torque. As we have shown in \S\ref{sec:trace},
the interference by other nearby embryos induces mixing of 
stream lines and restores diffusion across their corotation 
regions. 


\section{Discussion and Summary}
\label{sec:summary}
In the sequential accretion scenario, the formation of critical
mass cores (with $M_p > M_c \sim 10 M_\oplus$) prior to the 
severe depletion of gas in protostellar disks is a prerequisite
for the formation of gas giant planets.  In this paper, we 
present simulations to examine the condition for the formation
of critical mass cores.  

We confirm previous hypothesis that the core formation probability
is greatly enhanced by the type I migration of protoplanetary embryos.
The magnitude and direction of embryos' migration is determined by
their net differential Lindblad and corotation torque.  In the 
outer irradiated regions of protostellar disks, these torques lead 
to inward migration.  But in the inner viscously heated regions of 
the disk, unsaturated (full strength) corotation torque is stronger
than the differential Lindblad torque and it induces an outward migration.
However, the corotation torque is suppressed by saturation for both
massive ($> 10 M_\oplus$) and low-mass ($< 3 M_\oplus$) embryos.  
Nevertheless, super-Earth embryos undergo convergent migration towards
the transition radius ($r_t \sim$ a few AUs) which separates these 
two disk regions.

Here we present numerical simulations to show that the torque prescription
previously constructed by \citet{2010MNRAS.401.1950P, 2011MNRAS.410..293P}
for idealized single power-law $\Sigma_g$ and $S_g$ distribution can be 
generalized to more complex disk models.  This applicability is a reflection
that the embryo-disk torque is mostly applied to the proximity of the planets
orbit over length scales much shorter than that modifies the disk structure.

We show that when multiple embryos congregate near $r_t$, they interact with
each other through secular and resonant perturbations. In disks with modest
$\Sigma_g$ (or $\dot M$), the convergent speed is relatively slow. They would
capture each other on their mutual mean motion resonances if the resonant 
liberation time scale is shorter than that for the approaching embryos 
to cross the resonant width.  These planets form a convoy of resonant embryos.
The separation between the planets decreases with $\dot M$.  

In disks with sufficiently large $\dot M$, the embryos' separation becomes 
comparable or smaller than the total width of their mean motion resonance.
Interference between horseshoe stream lines around each planet enhances 
diffusion of gas through the corotation region and suppresses the saturation
of corotation resonances.  In this limit, embryos converge with overlapping 
orbits.  

Orbit crossing embryos undergo repeated close encounters.  Although some 
embryos are scattered outside the corotation zones, converging type I 
migration continually repatriate them back to the proximity of $r_t$.  
The trapped embryos continue to scatter each other until they undergo 
physical collision.  

The conditions for the collisional products to evolve into a super critical 
mass core are: 1) there is adequate time for collision to occur, 2) the 
physical collisions need to be mostly cohesive, and 3) the merger products 
must be retained.  The value of $r_t$ is also an increasing function of 
$\dot M$.  In disks with sufficiently large $\dot M$ for embryos to bypass
the resonant barrier, $r_t$ is at least a few AU where the local Keplerian 
velocity is smaller than the surface escape speed of super-critical mass 
cores.  Since the embryos' velocity dispersion is much smaller than their 
Keplerian velocity, their collisions are not sufficiently energetic to cause
any significant fragmentation.  

The collision time scale for the embryos is expected to be much longer than
the orbital period but shorter than the gas depletion time scale.  We used
an idealized model to simulate the consequence of a merger event on the
residual embryos.  Hydrodynamic simulations with a realistic collision
time scale is beyond the current computational feasibility, especially
for a systematic model parameter study.  Based on the results presented 
here, we will present a newly constructed HERMIT-Embryo code to simulate 
the evolution of multiple embryos in subsequent papers of this series.  
We will also replace the idealized composite power-law $\Sigma_g$ 
distribution with comprehensive evolving disk models.  


\section{Acknowledgement}
We thank Dr. Clement Baruteau for discussions and technical assistance. 
We also thank Prof. S. Ida for very useful conversations and an anonymous
referee for helpful suggestions in the presentation of this paper.
This work is supported by a grant from UC/Lab fee program.  
We acknowledge support from the LDRD program and IGPPS of Los Alamos National Laboratory.

\bibliographystyle{apj}
\bibliography{reference}

\end{document}